\documentclass[final,5p,times,twocolumn]{elsarticle}
\usepackage{subfigure}
\usepackage{amssymb}
\usepackage{lipsum}
\usepackage{amsthm}
\usepackage{graphicx}
\usepackage{amsmath}
\usepackage[colorlinks=true, linkcolor=blue, citecolor=blue, urlcolor=blue]{hyperref}

\journal{European Physical Journal C}

\begin{document}

\begin{frontmatter}

\title{Chaotic particle dynamics near a traversable wormhole throat}

\author[first]{Xin Zhao}
\ead{zhaox923@nuaa.edu.cn}
\author[first]{Xing-Kun Zhang}
\ead{zhangxk@nuaa.edu.cn}
\author[first,third]{Ya-Peng Hu}
\ead{huyp@nuaa.edu.cn}
\author[first,third]{Yu-Sen An\fnref{label1}}
\ead{anyusen@nuaa.edu.cn}
\fntext[label1]{Corresponding author}
\affiliation[first]{organization={Center for the Cross-disciplinary Research of Space Science and Quantum-technologies (CROSS-Q), College of Physics, Nanjing University of Aeronautics and Astronautics, 29 Jiangjun Road, Nanjing City, Jiangsu Province 211106, China}}
\affiliation[third]{organization={MIIT Key Laboratory of Aerospace Information Materials and Physics,  Nanjing University of Aeronautics and Astronautics, Nanjing, 210016, China}}

\begin{abstract}
This study investigates the nonlinear dynamics of a test particle near the traversable wormhole throat under an external harmonic potential. One-dimensional radial perturbation analysis shows that the particle is locally linearly stable at the equilibrium position. 
However, for two-dimensional and high-energy cases, the system exhibits a nonlinear response, leading to large-scale chaos. The analysis indicates that, if the particle is confined on one side of the wormhole, the Poincaré section 
will still retain Kolmogorov-Arnold-Moser (KAM) tori under extremely high-energy conditions, which is distinct from the chaos caused by the event horizon in the black hole. By studying another set of shape functions, the universality of this phase space structure is confirmed. This research clarifies the unique nonlinear dynamical mechanism of a traversable wormhole. It provides a new criterion, based on chaotic dynamics, for identifying black hole mimickers in strong-field astrophysical observations.

\end{abstract}

\begin{keyword}
traversable wormhole \sep particle motion \sep chaotic dynamics
\end{keyword}

\end{frontmatter}

\section{Introduction}
Over the past years, black holes have transcended their status as mere solutions of general relativity to become observationally verified astrophysical objects~\cite{LIGOScientific:2016aoc, EventHorizonTelescope:2019dse, EventHorizonTelescope:2022wkp}. A defining feature of a black hole is its event horizon, where its intense gravitational field governs the dynamics of the surrounding matter, including accretion flows, jets, and the orbital motions of compact objects. Within this context, the chaotic particle motion near the horizon has attracted considerable attention. Refs.~\cite{Hashimoto:2016dfz, Dalui:2018qqv} showed that a particle pulled by an external force develops chaos close to the horizon. The resulting Lyapunov exponent $\lambda$ is universally given by the surface gravity $\kappa$, independent of the form of the external force or the particle mass, and saturates the universal upper bound $\lambda \leq 2\pi T/\hbar$, conjectured by Maldacena, Shenker, and Stanford (MSS)~\cite{Maldacena:2015waa} from quantum field theory. This result establishes a connection between horizon geometry and chaos.

Subsequent works have extended this research in various directions, including possible violations of the MSS bound~\cite{Zhao:2018wkl, Lei:2020clg, Lei:2021koj, Gao:2022ybw, Gwak:2022xje, Yu:2023spr, Jeong:2023hom, Prihadi:2023tvr, Dutta:2024rta}, using chaotic particle motion as a probe of dark matter~\cite{Das:2025vja, Ali:2025ooh, Das:2025eiv}, assessing its impact on gravitational waveforms in extreme-mass-ratio inspirals~\cite{Das:2025eiv}, or incorporating quantum corrections~\cite{Bera:2021lgw, An:2025xmb}.

Given that all the aforementioned studies are conducted in spacetimes with event horizons, a natural and important question arises. Is this a unique feature for black hole event horizons, or could it also occur in other ultra-dense celestial objects without event horizons? The latter, as black hole mimickers, have attracted considerable attention in recent years.

Among various black hole mimickers~\cite{Morris:1988cz, Mazur:2001fv, Schunck:2003kk, Liebling:2012fv, Raposo:2018rjn, Cardoso:2019rvt, Mazza:2021rgq, Bambi:2025wjx}, traversable wormholes stand out due to their unique charm and have become a particularly notable candidate. Unlike black holes, traversable wormholes do not have an event horizon. They have a "throat" that connects two asymptotically flat (or AdS) regions. The absence of horizons means that, in principle, signals can propagate from one side of the wormhole to another side through the throat. In 1988, Morris and Thorne proposed a theoretical wormhole model that could be traversed by human beings~\cite{Morris:1988cz}. Since then, many publications have presented various new types of wormholes and related studies, such as Refs.~\cite{Visser:1989kh, Visser:2003yf, Lobo:2005us, Lobo:2009ip, Kanti:2011jz, Blazquez-Salcedo:2020czn, Konoplya:2021hsm, Kain:2023ore}.

In recent years, breakthroughs in gravitational wave astronomy have the ability to verify the distinction between black holes and exotic compact objects by analyzing oscillatory signals, tidal deformations, or echo characteristics~\cite{LIGOScientific:2016aoc, LIGOScientific:2017vwq, LIGOScientific:2018hze, LIGOScientific:2020aai, Ajith:2024mie}. However, based on the dynamical features of chaotic motion near the traversable wormhole throat, especially the fundamental differences between its chaotic behavior and that of black holes, systematic research in this regard has not yet been conducted.

Therefore, we investigate the nonlinear dynamical evolution of a test particle near the throat of an ultrastatic traversable wormhole under an external harmonic confining potential, and reveal its distinction from the chaotic behavior observed near the black hole horizon. Under analytical perturbations, our analysis shows that for radial motion, the test particle confined by an external harmonic potential is locally linearly stable at equilibrium positions. Interestingly, as the equilibrium position approaches the wormhole throat ($r_e \rightarrow r_0$), $\lambda^2 \rightarrow 0$. Thus, the particle dynamics outside the traversable wormhole are far more stable than the black hole case. This analytical feature stands in qualitative contrast to the typically unstable structures observed near the black hole horizon, which suggests that the chaotic motions outside the wormhole will also be less dramatic than in the black hole case, laying an important theoretical foundation for understanding the subsequent evolution of high-dimensional nonlinear dynamics.

The chaotic behavior can be demonstrated by analyzing the phase space of particle motion in two-dimensional space, where we introduce an additional angular direction. By solving the Hamiltonian canonical equations with the fourth-order Runge-Kutta (RK4) method and plotting Poincaré sections, we find that the two-dimensional system exhibits extensive chaos under high-energy excitation. This chaos originates from the nonlinear geometric coupling term induced by spacetime curvature. Crucially, numerical results confirm that even under extremely high-energy excitation, due to the presence of local linear stability mechanisms in the radial direction, partly KAM tori persist stubbornly in the core region of phase space near the equilibrium point. This phenomenon stands in sharp contrast to the widespread chaos observed in Schwarzschild black holes under the same confining potential~\cite{Hashimoto:2016dfz, Dalui:2018qqv}, which arises due to the event horizon. This discovery provides an effective criterion, based on nonlinear dynamical behavior, for identifying black hole mimickers in astrophysical observations in strong gravitational fields. 

The remainder of this article is organized as follows. Sec.\ref{Sec2} presents the background metric and conducts a one-dimensional radial perturbation instability analysis; Sec.\ref{Sec3} establishes a complete two-dimensional Hamiltonian canonical dynamical evolution framework and analyzes the geometric coupling mechanism; Sec.\ref{Sec4} presents the Poincaré sections and the maximum Lyapunov exponent; Sec.\ref{Sec5} presents the results for another wormhole shape function; Sec.\ref{Sec6} presents our main conclusions and gives an outlook.

\section{Dynamical Stability of the Particle Near the Wormhole Throat}\label{Sec2}
We consider particle motion in the background of a Morris-Thorne wormhole, a static, spherically symmetric spacetime. The metric is given by~\cite{Morris:1988cz}
\begin{equation}\label{metric}
\begin{aligned}
        ds^2=-e^{2\Phi(r)}dt^2+\frac{dr^2}{1-b(r)/r}+r^2 d\Omega^2~,
\end{aligned}
\end{equation}
where $\Phi(r)$ and $b(r)$ are the redshift and shape functions, respectively. The wormhole throat is at $r=r_0$, defined by $b(r_0)=r_0$.

The Lagrangian of a particle of mass $m$ moving only in the radial direction under an external potential $V(x^\mu)$ is given by
\begin{equation}\label{eqlag}
\begin{aligned}
    \mathcal{L}&=-m\sqrt{-g_{\mu\nu}\dot{x^\mu}\dot{x^\nu}}-V(r)\\
               &=-m\sqrt{e^{2\Phi(r)}-\frac{\dot{r}^2}{1-b(r)/r}}-V(r)~.
\end{aligned}
\end{equation}
where the dot in $\dot{r}$ is with respect to $t$. For small radial perturbations $\dot{r}\ll 1$, the Lagrangian for a particle can be expanded as
\begin{equation}
\begin{aligned}
    \frac{\mathcal{L}}{m}=\frac{g_{rr}(r)\dot{r}^2}{2\sqrt{-g_{tt}(r)}}-V_{eff}=\frac{e^{-\Phi(r)}\dot{r}^2}{2(1-b(r)/r)}-V_{eff}~.
\end{aligned}
\end{equation}

where
\begin{equation}
\begin{aligned}
    V_{eff}=\sqrt{-g_{tt}(r)}+\frac{V(r)}{m}=e^{\Phi(r)}+\frac{V(r)}{m}~.
\end{aligned}
\end{equation}

The effective potential $V_{eff}(r)$ depends on both the redshift function $\Phi(r)$ and the external potential $V(r)$. At the particle’s equilibrium position $r=r_e$, the derivative of the effective potential vanishes, $V'_{eff}(r_e)=0$, which leads to 
\begin{equation}
\begin{aligned}
    V'(r_e)=\frac{mg_{tt}'(r_e)}{2\sqrt{-g_{tt}(r_e)}}=-me^{\Phi(r_e)}\Phi'(r_e)~.
\end{aligned}
\end{equation}
Near the equilibrium position $r=r_e$, the Lagrangian can be expanded as
\begin{equation}\label{eqlaglage}
\begin{aligned}
    \frac{\mathcal{L}}{m}&\sim \frac{g_{rr}(r_e)}{2\sqrt{-g_{tt(r_e)}}}\left( \dot{r}^2+\lambda^2 (r-r_e)^2\right)\\
    &=\frac{e^{-\Phi(r_e)}}{2 (1-b(r_e)/r_e)}\left(\dot{r}^2+\lambda^2(r-r_e)^2\right)~,
\end{aligned}
\end{equation}
where
\begin{equation}\label{eqlambda}
\begin{aligned}
    \lambda^2&=-\frac{\sqrt{-g_{tt}(r)}}{g_{rr}(r)}V_{eff}''(r) \Big|_{r=r_e}\\
             &=-e^{\Phi(r)} \left(1 - \frac{b(r)}{r}\right) \left(e^{\Phi(r)}\Phi'(r)^2+e^{\Phi(r)}\Phi''(r) + \frac{V''(r)}{m}\right)\Big|_{r=r_e}~.
\end{aligned}
\end{equation}

The radial stability of the system near the equilibrium point depends on the sign of $\lambda^2$. If $\lambda^2>0$, the system is unstable. If $\lambda^2<0$, the system is stable. For the unstable case, the equation of motion takes the following form
\begin{equation}\label{eqddr}
    \ddot{r} =\lambda^2(r-r_e)~,
\end{equation}
the trajectory describing the particle in the radial direction can be solved as
\begin{equation}\label{eqr}
    r=r_e+Ae^{\lambda t}+Be^{-\lambda t}~.
\end{equation}
While for the stable case $\lambda^{2}<0$, the particle will oscillate around the equilibrium point. We mainly focus on ultrastatic wormholes ($\Phi(r)=0$). In this case, near the wormhole throat, Eq.~(\ref{eqlambda}) can be expanded as a series
\begin{equation}\label{lambdanew}
\begin{aligned}
    \lambda^2=-\frac{1-b'(r_0)}{r_0}\frac{V''(r_0)}{m}(r_e-r_0)+\mathcal{O}(r_e-r_0)^2~.
\end{aligned}
\end{equation}

It can be seen that near the wormhole throat, the leading term in the $\lambda^2$ series expansion is zero as a result of $b(r_{0})=r_{0}$, while the wormhole shape function and the external potential field jointly affect the higher-order terms. According to the flare-out condition for a traversable wormhole, geometrically, this requires that at the throat, $b'(r_0)<1$. This implies that $\mathcal{F}_0 \equiv (1 - b'(r_0))/r_0$ is strictly positive. Furthermore, since $r_{0}$ is the minimum size of the wormhole, its equilibrium position satisfies $r_e>r_0$. Therefore, the sign of $\lambda^2$ is determined entirely by the second derivative of the external potential field, $V''(r_0)$.

When the external potential is a conventional spatial confining potential (such as a one-dimensional harmonic oscillator potential with $V''>0$), Eq.~(\ref{lambdanew}) yields $\lambda^2<0$. This indicates that, under radial perturbations, the equilibrium orbits near the throat of an ultrastatic wormhole are linearly stable. This shows that the spacetime geometry of an ultrastatic wormhole, by itself, does not provide a nonlinear gravitational potential capable of inducing one-dimensional local chaos. As the equilibrium position approaches the throat ($r_e\rightarrow r_0$), $\lambda \rightarrow 0$. 

It is worth emphasizing that the one-dimensional radial instability analysis above stands in striking contrast to the dynamics of a particle near the black hole horizon. Consider a Schwarzschild black hole.
\begin{equation}\label{bhmetric}
\begin{aligned}
        ds^2=-f(r)dt^2+\frac{dr^2}{f(r)}+r^2 d\Omega^2~,
\end{aligned}
\end{equation}
with $f(r)=1-2M/r$, and the horizon is located at $f(r_h)=0$. According to the radial stability analysis above, substitute the Schwarzschild metric into Eq.~(\ref{eqlambda}), and expand it near the horizon to obtain
\begin{equation}\label{lambdabh}
\begin{aligned}
    \lambda^2=\frac{f'(r_h)^2}{4}+\mathcal{O}(r_e-r_0)^2=\kappa^2+\mathcal{O}(r_e-r_0)^2~.
\end{aligned}
\end{equation}
Note that here the leading term is non-zero for the black hole spacetime since $g_{tt}=g_{rr}^{-1}$, which is different from the Morris-Thorne wormhole. 

For a particle moving near the black hole horizon, its chaotic behavior is typically governed by the surface gravity, yielding a universal Lyapunov exponent $\lambda=\kappa$~\cite{Hashimoto:2016dfz, Dalui:2018qqv}. This single-particle result saturates the well-known MSS bound ($\lambda \le 2\pi T/\hbar$), originally proposed for many-body systems~\cite{Maldacena:2015waa}.\footnote{Note that, the MSS limit can sometimes be violated in single-particle mechanics under certain specialized setups~\cite{Zhao:2018wkl, Lei:2020clg, Lei:2021koj, Gao:2022ybw, Gwak:2022xje, Yu:2023spr, Jeong:2023hom, Prihadi:2023tvr, Dutta:2024rta}.} 
Thus, the black hole horizon can be seen as a nest of chaos. However, for the wormhole spacetime, as the perturbation is stable, there is no such chaos in the radial motion of massive geodesics. So from this aspect, wormhole spacetime is not as chaotic as black hole spacetime.\footnote{Since the temperature and thermodynamics of the wormhole are not so straightforward as the black hole due to the lack of the horizon, the discussion of the MSS chaos bound may remain obscure. However, there are indeed some formulations of the wormhole thermodynamics based on generalized first law \cite{Hayward:2009yw, Martin-Moruno:2009rpi}, where the wormhole temperature is still related to the surface gravity by $T=\frac{\kappa}{2\pi}$.}

Although one-dimensional linear analysis predicted the orbit's stability, the actual motion of astrophysical particles is not necessarily confined to a single direction. To explore whether there is truly nonlinear chaotic behavior in an ultrastatic wormhole and to reveal the differences between traversable wormholes and black holes at the nonlinear dynamical level, we will further investigate the particle motions in a two-dimensional plane. As shown in the subsequent section, the strong nonlinear $\theta-r$ coupling introduced by the spacetime geometry may lead to the emergence of chaotic phenomena in the high-energy region. The one-dimensional local stability near the throat, as demonstrated in this section, has become the theoretical basis for the persistence of the KAM tori in the high-energy chaotic region.

\section{Hamiltonian Analysis}\label{Sec3}
In this section, we investigate the phase space structure of particle motion near the wormhole throat using the Hamiltonian formulation. Due to the time-translational invariance of the static background spacetime, there exists a timelike vector $\chi^\alpha=(1,0,0,0)$ for an ultrastatic wormhole. The energy is given by $E=-\chi^\alpha p_\alpha=-p_t$, where $p_\alpha=(p_t,p_r,p_\theta,p_\phi)$ is the four-momentum vector. The covariant form of the dispersion relation for the particle reads
\begin{equation}
    g^{\mu\nu}p_\mu p_\nu=-e^{-2\Phi(r)}p_t^2+\frac{p_\theta ^2}{r^2}+p_r^2(1-\frac{b(r)}{r})=-m^2~,
\end{equation}
where $m$ is the mass of the particle. Without loss of generality, the particle is assumed to move in the $r-\theta$ plane, so $p_\phi=0$.\footnote{The motion in the $r-\theta$ plane is equivalent to the motion in the $r-\phi$ plane with $\theta=\frac{\pi}{2}$ due to rotational invariance.}

The energy of a massive particle takes two values, and we choose the positive sign for the outgoing particle in the external potential $V(r)$. Therefore, the explicit form of the particle's Hamiltonian $\mathcal{H} \equiv E$ is given by
\begin{equation}\label{ham}
    \mathcal{H} \equiv E=\frac{e^{\Phi(r)} \sqrt{p_\theta^2+(m^2+p_r^2)r^2-p_r^2\,r\,b(r)}}{r}+V(r,\theta)~.
\end{equation}

The Hamiltonian equations are derived in the standard way
\begin{equation}\label{eqdr}
    \dot{r}=\frac{\partial\mathcal{H}}{\partial p_r}
    =\frac{e^{\Phi} p_r \left( r - b(r) \right)}{\sqrt{ p_{\theta}^{2} - p_r^{2} \,r \,b(r) + \left( m^{2} + p_r^{2} \right) r^{2} }}~,
\end{equation}
\begin{equation}\label{eqtheta}
    \dot{\theta}=\frac{\partial\mathcal{H}}{\partial p_\theta}
    =\frac{e^{\Phi} p_{\theta}}{r \sqrt{p_{\theta}^{2} - p_{r}^{2}\, r\, b(r) + \left(m^{2} + p_{r}^{2}\right) r^{2}}}~,
\end{equation}
\begin{equation}\label{eqpr}
\begin{aligned}
   \dot{p_r}=-\frac{\partial\mathcal{H}}{\partial r}
    =\frac{e^{\Phi} \left( 2p_\theta^2-p_r^{2}\,r\,b(r)+p_r^{2}\,r^2\, b'(r) \right)}{2 r^2 \sqrt{ p_{\theta}^{2} - p_r^{2}\, r\, b(r) + \left( m^{2} + p_r^{2} \right) r^{2} }}\\
    - \frac{e^{\Phi} \sqrt{ p_{\theta}^{2} -  p_r^{2} \,r\,b(r) + \left( m^{2} + p_r^{2} \right) r^{2} }\Phi'}{r}-\frac{\partial V(r,\theta)}{\partial r}~,
    \end{aligned}
\end{equation}
\begin{equation}\label{eqptheta}
    \dot{p_\theta}=-\frac{\partial\mathcal{H}}{\partial \theta}=-\frac{\partial V(r,\theta)}{\partial \theta}~.
\end{equation}

\begin{figure*}[ht]
    \includegraphics[width=\textwidth]{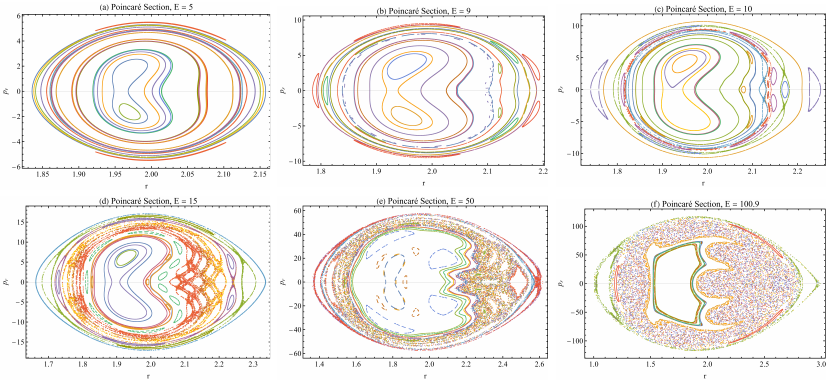}\hfill
    \caption{Poincaré sections of a massive particle for $b(r)=r_0^2/r$ at $E=$5, 9, 10, 15, 50, 100.9. As the energy increases, the phase space undergoes a transition from regular KAM tori to predominantly chaotic motion, with surviving KAM tori persisting amidst the irregular scatter.}
    \label{psm1}
\end{figure*}

From the derived Hamiltonian equations (\ref{eqdr})-(\ref{eqptheta}) above, it is clear that the strong nonlinear coupling mechanism triggered by the curved spacetime geometry is evident. Due to this composite nonlinear structure, the radial motion ($r, p_r$) and the angular motion ($\theta, p_\theta$) cannot be decoupled analytically. In previous studies regarding black hole spacetime \cite{Hashimoto:2016dfz, Dalui:2018qqv}, when the total energy $E$ of the system is low, the two directions are approximately decoupled, resulting in regular quasi-periodic motion of the particle; however, as the total energy increases, the particle motions will gradually transit to chaotic behavior. We will show in the following that these patterns will persist in wormhole spacetime. 

Although this two-dimensional Hamiltonian system exhibits extremely strong nonlinearity at high energy, it is worth noting that even when high-energy nonlinear coupling leads to extensive chaos, the radial local stability mechanism ensures that the central region of the phase space may still stubbornly retain the KAM tori that do not break. This analysis will be directly confirmed in the subsequent investigation of the Poincaré section.

\section{Poincaré Section and Maximum Lyapunov Exponent}\label{Sec4}
To reveal the specific dynamical evolution in phase space, we must specify the shape function of the wormhole and the external potential. We adopt an ultrastatic traversable wormhole background, for which the redshift function satisfies $\Phi(r)=0$. Drawing on previous studies of traversable wormholes, we choose the shape function $b(r)=r_0^2/r$~\cite{Ellis:1973yv, Muller:2008zza, Ohgami:2015nra, Tsukamoto:2016qro}. Additionally, to effectively confine a particle near the throat, we introduce a two-dimensional isotropic local harmonic potential, given by:
\begin{equation}\label{eV}
    V(r,\theta)=\frac{1}{2}m\omega_r^2(r-r_e)^2+\frac{1}{2}m\omega_\theta^2\theta^2~.
\end{equation}
Here, $\omega_r$ and $\omega_\theta$ represent the particle's binding strengths in the radial and angular directions, respectively, while $r_e$ is the particle's equilibrium position when it moves purely radially.

By redefining $E=E/m$, $p_{r}=p_{r}/m$ and $p_{\theta}=p_{\theta}/m$, the mass $m$ can be set to unity without loss of generality. To investigate the transition of the system from quasi-periodic integrable motion to nonlinear chaotic motion in depth, we set the following parameters in the numerical calculation.
\begin{equation}
\begin{aligned}
    &m=1~,\quad r_0=1~, \quad r_e=2~, \\ &  \omega_r=10\sqrt{2}~,\quad \omega_\theta=10\sqrt{2}~.
\end{aligned}
\end{equation}

Because the particle is confined to one side of the wormhole, we expect that it will not pass through the throat of the wormhole. Thus, its energy cannot exceed the critical energy imposed by the external potential. The critical condition is deduced by demanding that when the particle approaches the throat $r=r_{0}$ from direction $\theta=0$, its momentum decreases to zero $p_r=p_{\theta}=0$. Once a particle's initial energy exceeds this barrier, it can traverse the throat to the other side of the wormhole. For the convenience of comparison with the black hole case, we do not consider this situation during this work. By substituting the critical condition into the Hamiltonian Eq.~(\ref{ham}), we can calculate that the critical energy is $E_c=101$. To comprehensively observe the transition from integrable to chaotic behavior in phase space, we selected six representative energies: $E=5$, $E=9$, $E=10$, $E=15$, $E=50$, and $E=100.9$, which gradually approach the critical energy.

We integrated the previously established two-dimensional Hamiltonian equations (\ref{eqdr})-(\ref{eqptheta}) using the RK4 method. The initial states $(r_i, \theta_i, p_{ri})$ were randomly sampled for fixed energy, while the angular momentum $p_{\theta i}$ was obtained from the Hamiltonian equation in Eq.~(\ref{ham}). In order to get the Poincaré section for the $(r, p_r)$ plane, the phase space evolution ($t \in [0,10^4]$) was projected onto a two-dimensional hypersurface where $\theta=0$ with $p_\theta>0$.

Fig.\ref{psm1} shows the evolution of the Poincaré section in phase space near the wormhole throat for different energies. In each subgraph, points of the same color represent the same initial value $(r_i, \theta_i, p_{ri},p_{\theta i})$.\footnote{Points of the same color in different subgraphs do not share this meaning.} At lower energy, such as $E=5$ (Fig.\ref{psm1}(a)), the phase space is dominated by a series of smooth, nested, closed curves, namely the regular KAM tori. At this energy, the particle's kinetic energy and angular momentum excitation are extremely low, and the geometric nonlinear coupling term is strongly suppressed. The particle undergoes quasi-periodic, regular oscillations near the equilibrium position $r_e=2$.

As the energy increases, the phase space develops a distinctive structure at the energies $E=$5, 9, 10, 15, 50, 100.9 (Fig.\ref{psm1}(b)-(f)). Within the phase space, partial KAM tori start to break, and the system begins to exhibit chaotic structure. It is clear that as the energy increases, the chaotic regions in the Poincaré section also grow larger. Since we adopted the same evolution time for different initial values, this implies that the Lyapunov exponent, which characterizes the chaos of the system, also increases. This is consistent with the subsequent numerical results for the maximum Lyapunov exponent.

When the energy approaches the critical value, $E=100.9$ (Fig.\ref{psm1}(f)), the Poincaré section shows that, due to the injection of extremely high initial energy, the previously smooth and numerous closed KAM tori are completely torn apart and melt into diffuse, chaotic points in phase space. However, the highlight of the entire nonlinear dynamical system is that even when the KAM tori melt into chaotic points under high energy, the central region of the phase space still harbors a large number of dense local KAM tori that stubbornly survive.

The persistence of these KAM tori precisely matches and validates our previous analytical prediction from one-dimensional linear perturbation theory. In the numerical experiments, we fixed the equilibrium position of the external potential at $r_e=2$. The Poincaré section (Fig.\ref{psm1}) clearly shows that, whether for $E=$5, 9, 10, 15, 50 or $E=100.9$, a large number of unbroken closed KAM tori always exist in the central region of the phase space, that is, in the local vicinity of the equilibrium point $r_e=2$. According to the previous analytical perturbation analysis, when the particle undergoes small oscillations near the equilibrium point $r_e$, the eigenvalue of the one-dimensional radial perturbation satisfies the stability condition ($\lambda^2<0$). Therefore, the KAM tori persist around the equilibrium point $r_e=2$ at high energy, providing numerical evidence for the previous stability analysis. It qualitatively indicates that although high energy excites widespread chaos, the local stability mechanism near the equilibrium point remains effective, forming KAM tori that resist chaos.

The evolution from near-integrable KAM tori at low energy to developed chaos at higher energy shows that increasing energy drives the system away from integrability while largely preserving regular structures.
This partial preservation of regularity contrasts with the integrability-to-chaos transition as observed in particle motion near the black hole horizon \cite{Hashimoto:2016dfz, Dalui:2018qqv}.

Although the Poincaré section provides qualitative insight into the structure of phase space, the maximum Lyapunov exponent is required to rigorously and quantitatively distinguish between chaotic and regular dynamics. The maximum Lyapunov exponent $\lambda_{\text{m}}$ measures the asymptotic rate at which two extremely close orbits in phase space diverge exponentially over time. Its formal definition is:
\begin{align}
\lambda_{\text{m}}=\lim_{t\to \infty}\frac{1}{t} \ln \frac{d_i}{d_0}~.
\end{align}
Among them, $d_0$ and $d_i$ represent, respectively, the distance between the two trajectories in phase space at the initial and instantaneous times.

\begin{figure}[h]
    \centering
    \includegraphics[width=0.45\textwidth]{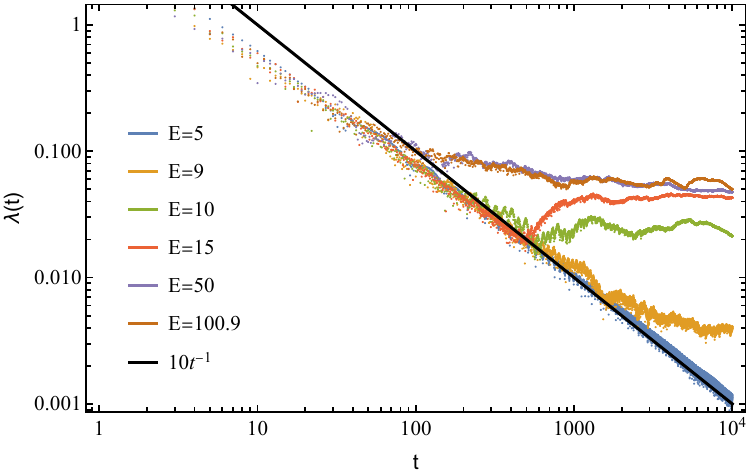}
    \caption{When $E=$5, 9, 10, 15, 50, 100.9, the log-log graph shows the evolution of the maximum Lyapunov exponent in the corresponding Poincaré section (Fig.\ref{psm1}) over time, $t \in [0,10^4]$. When $t=10^4$, $\lambda_{\text{m}}=$[0.001, 0.004, 0.021, 0.043, 0.047, 0.050], corresponding to $E=$5, 9, 10, 15, 50, 100.9 from left to right.}
    \label{MLE1}
\end{figure}

In Fig.\ref{MLE1}, we present a log-log plot of the evolution of $\lambda_{\text{m}}$ over time at different energies ($E=$5, 9, 10, 15, 50, 100.9). In the long-time limit, these curves exhibit remarkably different dynamics.

\begin{figure*}[ht]
    \centering
    \includegraphics[width=\textwidth]{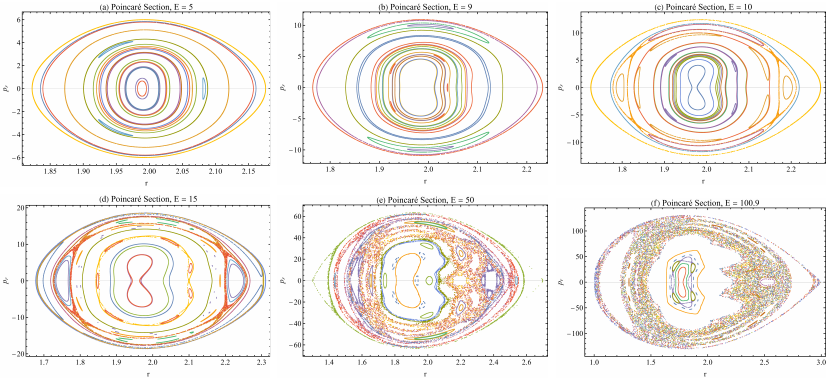}\hfill
    \caption{Poincaré sections of a massive particle for $b(r)=re^{-(r-r_0)}$ at $E=$5, 9, 10, 15, 50, 100.9. As the energy increases, the phase space transitions from regular KAM tori to predominantly chaotic motion, with surviving KAM tori persisting amid the irregular scatter.}
    \label{psm2}
\end{figure*}

For the regular KAM tori orbits, the maximum Lyapunov exponent over time exhibits a power-law decay and follows the scaling law $\lambda(t) \propto t^{-1}$ (corresponding to the black line). This clear power-law behavior indicates that the distance between two orbits in phase space fails to grow exponentially with time, thereby providing evidence for the regular quasi-periodic motion of the KAM tori.

By contrast, for orbits originating from the chaotic region, the value of $\lambda_{\text{m}}$ stops declining in the later stage of evolution and converges to a definite, non-zero plateau, $\lambda_{\text{m}}\approx\text{const}>0$. This remarkable plateau clearly confirms high sensitivity of the system to initial conditions, thereby quantitatively corroborating the chaotic nature observed in the Poincaré section.

It is worth noting that the maximum Lyapunov exponent also increases with energy, consistent with the qualitative results from the Poincaré section. However, as shown in Fig.\ref{MLE1}, at high energy, the maximum Lyapunov exponent of the system should saturate at a certain value.

In conclusion, the close agreement between the qualitative Poincaré section diagrams and the quantitative maximum Lyapunov exponent scaling characteristics strongly confirms the rich nonlinear behavior of the particle motion near the throat of the traversable wormhole.

\section{Generality Under Another Shape Function}\label{Sec5}
To ensure that the nonlinear dynamical behavior discovered in the previous section under a specific shape function 
remains general. This section introduces a new type of traversable wormhole shape function $b(r)=re^{-(r-r_0)}$ that is studied in Refs.~\cite{Samanta:2018hbw, Godani:2019kgy, Tangphati:2020mir, Dutta:2024luw}. We expect to verify the universality of the chaotic evolution of this strong-gravity system through comparative studies.

Fig.\ref{psm2} shows the particle motion Poincaré section calculated based on this different shape function under the same energy choices. It is clear that the phase space exhibits extremely high consistency. Overall, the systems all went through a process from regular KAM tori to the coexistence of chaos and KAM tori. Although the local phase orbits undergo deformation due to the details of the shape function and the difference in the initial value, the overall phenomenon still presents a typical coexistence of strong chaos and KAM tori.

\begin{figure}[h]
    \centering
    \includegraphics[width=0.45\textwidth]{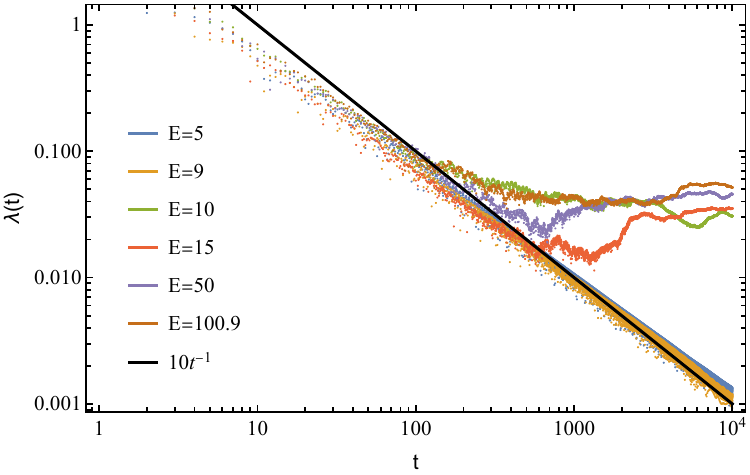}
    \caption{When $E=$5, 9, 10, 15, 50, 100.9, the log-log graph shows the evolution of the maximum Lyapunov exponent in the corresponding Poincaré section (Fig.\ref{psm2}) over time, $t \in [0,10^4]$. When $t=10^4$, $\lambda_{\text{m}}=$[0.001, 0.001, 0.031, 0.035, 0.046, 0.052], corresponding to $E=$5, 9, 10, 15, 50, 100.9 from left to right.}
    \label{MLE2}
\end{figure}
To provide quantitative evidence, Fig.\ref{MLE2} also presents the log-log plot of the maximum Lyapunov exponent over time for different energies under this shape function. The numerical results perfectly reproduce the dynamical characteristics. At high energies, the curves, starting from initial values in the chaotic region, rapidly converge to a plateau after a short decline, confirming strong sensitivity of the system to initial conditions. By contrast, at low energies, the curves, starting from the regular KAM tori, show a downward-sloping line with a slope of approximately "$-1$" on the log-log axis (corresponding to the black line), indicating that the distance between adjacent orbits within this local region grows only linearly.

Notably, our numerical results show that under two different shape functions ($r_0^2/r$ and $re^{-(r-r_0)}$), the maximum Lyapunov exponents of the testing particle consistently trend to the same saturation upper bound in the high-energy limit. This quantitative coincidence suggests that the chaotic saturation behavior near a traversable wormhole throat is bounded by an intrinsic ceiling.

The comparison results in this section strongly demonstrate that the nonlinear dynamic behaviors of the test particle near the throat of the traversable wormhole, as well as its transition to a large-scale chaotic state from KAM tori, do not depend on the specific shape function and exhibit physical universality.

\section{Conclusion and Discussion}\label{Sec6}
This paper systematically investigates the nonlinear dynamics of a test particle near the throat of the traversable wormhole. By solving the Hamiltonian equations of motion, we focus on the evolution of the system's phase space across different energy regimes ($E=$5, 9, 10, 15, 50, 100.9). Firstly, by plotting the Poincaré section, we find that the system exhibits rich nonlinear phenomena. The phase space does not flow entirely toward a completely non-integrable, extreme chaotic regime but instead exhibits a structure in which large-scale chaos and local regular KAM tori coexist. Secondly, we quantitatively compute the time evolution of the maximum Lyapunov exponent. The log-log graph shows that the Lyapunov exponents of chaotic orbits rapidly converge to the positive constant plateau ($\lambda_{\text{max}} \approx \text{const}>0$), confirming the exponential divergence characteristic of chaotic behavior; by contrast, regular KAM tori exhibit a maximum Lyapunov exponent that decays approximately as $t^{-1}$, indicating approximately integrable motion. In addition, the maximum Lyapunov exponent increases with energy and eventually reaches an upper limit. Finally, by introducing different wormhole shape functions $b(r)$ for comparison and verification, we find that the emergence of chaos and the asymptotic scaling behavior of the maximum Lyapunov exponent remain highly consistent, both qualitatively and quantitatively. This demonstrates that the chaotic dynamical feature does not depend on the specific details of the traversable wormhole shape functions.

The difference in the chaotic motion characteristics of the test particle near the throat of the traversable wormhole from those near the event horizon of the black hole~\cite{Hashimoto:2016dfz, Dalui:2018qqv} provides a new criterion based on chaotic dynamics for identifying black hole mimickers in strong-field astrophysical observations.

In future work, the test particle can be extended to massless photons, and chaotic light tracing near the wormhole throat can be studied. This will directly relate to the wormhole's shadow profile and provide potential theoretical criteria for future radio telescope observations. Alternatively, we only consider the particle motion on one side of the traversable wormhole in this work, a comprehensive analysis regarding the particle that can traverse the wormhole is also interesting to pursue in future work. Moreover, in this work, we only consider the symmetric wormhole, the analysis can also be generalized to the asymmetric wormhole configuration \cite{Ou:2021efv, Huang:2023yqd}. 
\section*{Acknowledgements}
We are grateful for the useful discussions with our group members. This work is supported by the National Natural Science Foundation of China (NSFC) under Grants No.12405066 and No.12175105. YSA is also supported by the Natural Science Foundation of Jiangsu Province under Grant No. BK20241376 and Fundamental Research Funds for the Central Universities. 

\appendix

\bibliographystyle{elsarticle-num} 
\biboptions{sort&compress}
\bibliography{msc}

\end{document}